**Bioelectrical interfaces beyond excitable cells: cancer, aging, and gene expression modulation**

*Paolo Cadinu,[1,2] Matthew Burgess,[3] Catarina Franco Jones,[4,5] Marzia Iarossi,[6] Manuel Schröter,[7] Nako Nakatsuka,[8] Mustafa B. A. Djamgoz,[9] Gil Gonçalves,[10,11] Sidahmed Abayzeed,[12] Paola Sanjuán-Alberte,[13,14] Paula M. Mendes,[15] Frankie J. Rawson,[16] Malavika Nair,[3] Michael Levin,[17,18] Rosalia Moreddu[4,5]**

[1]Program in Cellular and Molecular Medicine, Boston Children's Hospital, Boston, MA, USA
[2]Department of Microbiology, Blavatnik Institute, Harvard Medical School, Boston, MA, USA
[3]Institute of Biomedical Engineering, Department of Engineering Science, University of Oxford,
Oxford, UK
[4]School of Electronics and Computer Science, University of Southampton, Southampton, SO17 1BJ, UK
[5]Institute for Life Sciences, University of Southampton, Southampton, SO17 1BJ, UK
[6]Department of Biomedical Engineering, Technion Israel Institute of Technology, Haifa, Israel
[7]ETH Zurich, Department of Biosystems Science and Engineering, Basel, Switzerland
[8]Laboratory of Chemical Nanotechnology, Neuro-X Institute, EPFL, 1202 Genève, Switzerland
[9]Department of Life Sciences, Imperial College London, UK
[10]Centre for Mechanical Technology and Automation (TEMA), Department of Mechanical Engineering, University of Aveiro, 3810-193 Aveiro, Portugal
[11]Intelligent Systems Associate Laboratory (LASI), 4800-058 Guimarães, Portugal
[12]Faculty of Engineering, University of Nottingham, Nottingham, UK
[13]Department of Bioengineering and Institute for Bioengineering and Biosciences, Instituto Superior Técnico, University of Lisbon, Lisbon, Portugal
[14]Associate Laboratory i4HB, Institute for Health and Bioeconomy, Instituto Superior Técnico,
Universidade de Lisboa, Lisbon, Portugal
[15]School of Chemical Engineering, University of Birmingham, Birmingham, UK
[16]Bioelectronics Laboratory, Division of Regenerative Medicine and Cellular Therapies, School of Pharmacy, Biodiscovery Institute, University of Nottingham, Nottingham, UK
[17]Allen Discovery Centre at Tufts University, Tufts University, Medford, MA, USA
[18]Wyss Institute for Bioinspired Engineering at Harvard University, Boston, MA, USA
*Correspondence: r.moreddu@soton.ac.uk

## Abstract

The investigation of biological conductivity has evolved from its classical foundation based on ionic fluxes underpinning cardiac and neuronal excitability to a multifaceted regulator of cellular physiology. Traditional approaches for probing electrical events in living matter focused largely on action potentials recording. However, bioelectricity in non-excitable cells governs key phenomena, including developmental patterning, tissue homeostasis, and disease progression. Pioneering studies implicated endogenous bioelectrics in many aspects of morphogenesis, wound healing, regeneration, and cancer. Early findings laid the groundwork for viewing bioelectricity as a means to influence cell fate, cell cycle progression, differentiation, and senescence. More recently, spatial variations in membrane potential within tumor microenvironments were found to correlate with metastatic potential. In parallel, substantial breakthroughs have been achieved in designing advanced bioelectrical interfaces for the study of neuronal networks and cardiac function. This perspective bridges the engineering and biological domains by examining how such technologies might enable new insights into non-excitable cell electrical events at different scales of operation to ultimately manipulate cellular pathways in cancer reprogramming, anti-aging interventions, and gene expression modulation.

## 1. Introduction

Bioelectricity has traditionally been dominated by the exploration of excitable cells, most notably neurons and cardiomyocytes. In these cells, electrical events result from rapid ionic fluxes that give rise to action potentials, implicated in transmitting information and coordinating function, making them ideal systems to probe the principles of electrical signaling in biology.[1] Yet, a growing body of evidence has shown that bioelectricity is neither exclusive to excitable cells nor confined to fast signaling events.[2] Instead, it constitutes a fundamental layer of biological regulation, influencing processes like tissue patterning, morphogenesis, regeneration, gene expression, and disease progression.[2,3] Non-excitable cells exhibit rich electrical landscapes that can dictate fate decisions and even modulate organismal aging[4]. Bioelectronic devices have been instrumental in decoding the principles of cellular electrophysiology.[5] Microelectrode arrays,[6] nanoscale probes,[7] and quantum-enabled sensors[8] now allow precise interrogation of ionic currents, membrane potentials, and electrochemical signaling across scales[5]. Such tools are starting to be adapted to capture the lower-frequency and smaller-magnitude fluctuations characteristic of non-excitable cells, as well as high-frequency events in some non-excitable tissues.[3]

This shift presents both technical challenges and innovation opportunities. On one hand, it requires a rethink in sensitivity, stability, and biocompatibility to record and modulate electrical phenomena that were previously overlooked[9,10]. On the other, it opens the route to applying bioelectrical principles to address a series of biological and medical challenges[2]. In cancer, bioelectric signals are emerging as powerful biomarkers and potential therapeutic targets[11]. Tumor cells often display depolarized resting membrane potentials compared with their healthy counterparts, a state that correlates with enhanced proliferation, invasiveness, and metastatic potential[3]. Electrical modulation of ion channel activity has been shown to reprogram malignant signaling pathways[12], suggesting that bioelectronics could complement pharmacology, giving rise to the emerging field of electroceuticals[8]. Similarly, in the context of aging, depolarization of resting potentials is now recognized as a hallmark of senescence[4]. By restoring hyperpolarized states, it may be possible to slow or even reverse age-associated decline at the cellular level, opening new pathways in anti-aging interventions. Perhaps most transformative is the realization that membrane potential dynamics directly regulate gene expression.[13] Voltage changes influence calcium flux, activate transcription factors, and remodel chromatin, orchestrating transcriptional programs without the need for genomic manipulation[13,14]. The bioelectrical regulatory layer introduces new strategies for cellular reprogramming to achieve reversible control over lineage specification and differentiation[14].

Bioelectronics represents a powerful tool to interrogate and manipulate biology. Moving beyond excitable tissues, we can begin to comprehensively address how electric signals shape life, deepening our understanding of fundamental physiology and enabling novel interventional approaches. Dissecting new electrical communication principles in living systems might also enable the development of new bioinspired systems[15]. This perspective provides an overview of how bioelectronic devices could be deployed to modulate cancer, aging, and gene expression.

## 2. From quantum signaling to cellular network events

Bioelectronic devices for electrophysiology transformed our ability to probe and modulate cellular functions in excitable cells and tissues.[6] **Figure 1** displays selected current technologies operating at different spatial scales, and how these enable to access different biological phenomena. At the quantum level (sub-nanometer), quantum bioelectronics controls electron tunneling mechanisms behind all bioelectric events[8]. At the molecular scale (~1-10 nm), nanoswitches translate quantum effects into controllable molecular interactions[16]. At the subcellular scale (~10-100 nm), nanopores and nanotweezers enable precision biosensing and manipulation of biomolecules influenced by electrical states[7]. At the cellular and tissue level (~100 nm-100 μm), micro and nanoelectrodes record and stimulate cells[6], cell monolayers[17], and tissues[18]. The following sections presents how state-of-the-art approaches serve diverse functions and complement each other in interrogating and modulating electricity in living matter. Bioelectronic interfaces for non-excitable systems can be broadly grouped into (i) sensing platforms that map endogenous electrical and electrochemical variables (e.g., $V_{mem}$, ionic currents, impedance, redox/pH), and (ii) actuation platforms that deliver controlled stimulation, currents, or fields to impose defined electrical states. Because many device classes can be operated in both modes, here we discuss each technology in terms of its read (measurement) and write (stimulation) capabilities across length scales, and we summarize practical performance alongside charge-delivery constraints. **Table 1** compares performance and applications across the presented devices.

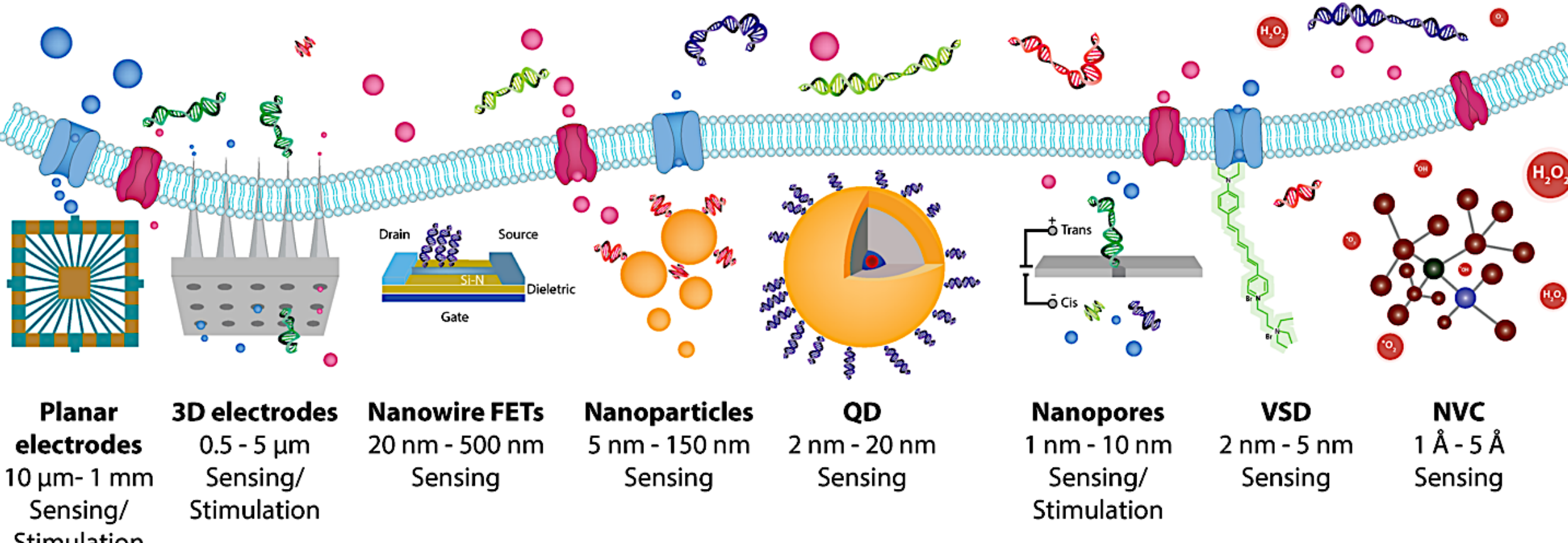

**Figure 1. Selected state-of-the-art bioelectronic technologies to probe and control biological activity across scales.** Schematics of representative platforms arranged from the quantum to the tissue level. At the atomic and molecular scale, quantum probes and electrically activated nanoswitches regulate charge transfer and interactions. Nanopores and nanotweezers enable single-molecule analysis and subcellular sampling. At the cellular and multicellular scale, planar and 3D nanoelectrodes provide access to extracellular and intracellular voltage dynamics. Quantum dots and nanoparticles offer optical, electrochemical, and therapeutic functionalities.

## 2.1. Micro and nanoelectrodes

Micro and nanoelectrodes have scaled how we access and manipulate electrical information in excitable cells, overcoming some limitations of patch-clamp technology.[6] Complementary metal-oxide-semiconductor (CMOS)-based high-density microelectrode arrays (HD-MEAs) now enable extracellular recording from thousands of sites simultaneously, capturing spatial electric patterns in neural and cardiac tissues with sub-cellular resolution[6]. Planar electrode architectures, however, often suffer from weak cell-electrode coupling and limited access to intracellular dynamics[6]. This limitation has motivated the development of vertical 3D nanoelectrodes,[19] which enhance cell-electrode interaction and, in some cases, breach the membrane to probe intracellular voltages or deliver direct electrical stimuli[19]. Meanwhile, flexible and stretchable electrodes fabricated from elastomeric substrates or soft conductive materials have enabled integration with mechanically active tissues, including vasculature, epithelium, and organoids.[18] Among these, conductive polymers like poly(3,4-ethylenedioxythiophene) polystyrene sulfonate (PEDOT:PSS) are considered particularly promising due to their volumetric capacitance, biocompatibility, and ability to form low-impedance contacts[20]. In oncology, such interfaces could be embedded in peritumoral or stromal compartments to monitor ionic gradients associated with immune evasion, stromal remodeling, or epithelial–mesenchymal transition (EMT), bioelectrical dimensions of tumor progression that remain underexplored[9, 11, 12]. Carbon-based materials, primarily graphene, carbon nanotubes (CNTs), and their composites, offer the additional advantages of high conductivity, electrochemical stability, mechanical strength, and optical transparency,[21] potentially enabling to combine electrophysiology with optical mapping to probe mitochondrial dynamics, cytoskeletal remodeling or chromatin architecture. CNTs can be vertically patterned to enhance cell-material adhesion and support subthreshold stimulation which could allow modulating signaling pathways without triggering full depolarization[21]. In the context of cancer, electrical stimulation via microelectrodes may be used to rewire malignant signaling without chemical intervention[3]. A key constraint for electrical modulation is the charge-injection capacity (CIC) of the electrode: for a given waveform, the delivered charge per phase Q over geometric electrode area A must remain within the reversible electrochemical limits of the material to minimize faradaic reactions and pH shifts at the interface. In established stimulation protocols, charge density is commonly discussed together with pulse parameters (e.g., via the Shannon model) and the requirement for charge-balanced waveforms.[22, 23] These considerations become especially important for non-excitable-cell studies, where modulation may use lower amplitudes but much longer durations (minutes-hours), so electrochemical by-products can dominate unless stimulation is either charge-balanced or delivered through electrochemically buffered configurations using capacitive coupling, salt bridges, or redox-stable electrode chemistries.[23, 24]

## 2.2. Molecular nanoswitches

Electrically responsive nanomaterials have been shown to modulate a variety of molecular processes. Their noninvasive nature, high spatiotemporal control, and reversible induction allows to influence cellular adhesion, release, alignment, polarization, migration, proliferation, and differentiation,[25] providing new opportunities to elucidate complex processes, from surface charge,[26] surface ligand conformations,[16] and mechanical properties[27]. Cell control through substrate manipulation has been mediated by conductive polymer-based scaffolds, peptide surfaces and piezoelectric materials[16]. For instance, changes in hydrophilicity induced by conductive polymer-based scaffolds have been utilized for capturing and releasing cancer cells.[28] Although nonspecific interactions can be used to regulate cell behavior, achieving more precise control is possible by specifically targeting the modulation of interactions between material surfaces and cells. Cell attachment to native extracellular matrices (ECMs) is orchestrated by cell-adhesion proteins such as fibronectin, collagen, and laminin, which bind selectively to receptors, such as integrins, on the cell surface.[16] The cell-ECM interaction is crucial for bidirectional signal transduction, guiding cell proliferation, spreading, and differentiation.[2] Progress has been made in mimicking the native ECM at the protein scale. Surface electrical potentials generated by piezoelectric materials have been shown to induce conformational changes in adsorbed fibronectin, promoting either cell adhesion or proliferation depending on the membrane potential.[2] Piezoelectric materials have also been used to polarize macrophages, accelerating wound recovery.[29] Stimuli-responsive materials have also been deployed in 3D printing, enabling structural or functional transformations over time to achieve "4D printing".[30] Various stimuli such as temperature and light have been explored in this context, but the full potential of electroresponsive material in manipulating cellular behavior remains underexplored. Current efforts have largely focused on building electroactive scaffolds, often coupled to dynamic nanotopography, to bias lineage commitment; for example, electrochemical switching of nanostructured polypyrrole interfaces provides coupled electrical and structural cues that can direct stem-cell differentiation, including osteogenic outcomes.[31] These advancements might bring us closer to creating dynamic scaffolds with organizational features and hierarchical architectures that mimic native tissues. Further developments are also necessary in the areas of multi-responsiveness and bidirectional actuation at the bio-interface. Advancing these aspects will, for instance, enhance the development of dynamic systems that better replicate the natural feedback mechanisms between cells and the ECM.

### 2.3. Nanopores and nanotweezers

Nanopores are increasingly recognized as powerful tools in bioelectricity, enabling mapping of electrical and electrochemical signals in living systems with molecular-level resolution.[7] Nanopipettes are tapered glass capillaries with nanoscale apertures (10–100s of nm) and serve as key sensing elements in Scanning Ion-Conductance Microscopy (SICM) and Scanning Electrochemical Microscopy (SECM), providing real-time insights into ion flow, action potentials, membrane potential fluctuations, and ion channel dynamics.[7] SICM integrates electrophysiological measurements with high-resolution topographic imaging of living cells under physiological conditions.[32] In addition to signal mapping, nanopipettes can function as biosensors. When configured as nanopores, transient ionic current fluctuations induced by biomolecule translocation enable single-molecule detection of DNA, RNA, peptides, and proteins.[33] Their functionality is further enhanced through chemical modification, improving molecular selectivity, transport control, and signal fidelity.[33] Examples include the use of DNA origami spheres to trap proteins via electroosmotic effects,[34] and DNA aptamers that bind specific amino acid motifs, slowing peptide translocation for improved resolution. For small-molecule sensing, aptamer-modified nanopipettes detect neurochemicals like serotonin and dopamine by undergoing voltage-sensitive conformational changes that modulate local charge distributions.[35] Beyond biosensing, nanopores have been demonstrated as tools for minimally invasive cellular nanobiopsy, allowing extraction of sub-picolitre volumes of cytoplasmic content from live cells with spatial and temporal resolution.[36] Nanopipette mounted on SICM were optically guided to penetrate a single cell, aspirate ~50 fL of cytosol containing mRNA or mitochondria, and re-eject the material for RNA sequencing via electrowetting.[37] Dual-barrel nanopipettes extend to subcellular resolution and longitudinal sampling, for example tracking transcriptional responses in glioblastoma cells before and after chemotherapy.[38] Unlike suction-based methods, nanotweezers operate via dielectrophoresis (DEP) and do not require fluid extraction. Two pyrolytic carbon electrodes are separated by an insulating septum and fabricated at the tip of a dual-barrel quartz pipette.[39] When an alternating electric field is applied, intense local DEP forces accumulate target biomolecules at the pipette apex, reducing cellular disruption, improving viability, and enabling repeated sampling. Applications include subcellular profiling of adrenergic receptor compartmentalization in cardiomyocytes, analysis of gene expression heterogeneity among mitochondria in primary neurons, and longitudinal molecular tracking in cultured breast cancer cells.[40]

**Table 1. Comparative overview of bioelectronic interfaces for non-excitable cells (sensing and modulation).**

| Platform class | Sensing readouts | Modulation modes | Spatial and temporal scales | Advantages | Limitations | Ref. |
|---|---|---|---|---|---|---|
| Planar or HD microelectrode arrays | Extracellular potentials, Slow field impedance changes | Local current injection, waveform control, multiplexed stimulation | ~10-100 μm; ms-hours (depending on readout) | High throughput, spatial patterns | Weak cell-electrode coupling for small or slow signals, drift, electrochemical limits for long stimulation | 6, 23 |
| 3D nanoelectrodes and nanopillars | Improved coupling, intracellular-like access | Intracellular stimulation (device-dependent) | ~10-100 nm features; ms-min | Higher coupling, potential intracellular access without full patch clamp | Fabrication complexity, variability, membrane damage risk, scaling challenges | 19 |
| Organic electrochemical or electrolyte-gated transistors | Amplified local potentials, mixed ionic-electronic coupling | Low-voltage coupling, potentially longer duration interfacing | ~10-100 μm; ms-min | High transconductance, soft conformal interfaces | Materials stability, encapsulation, standardization | 17, 18 |
| Electroactive switchable surfaces | Indirect: adhesion/traction/phenotype as functional readout | Voltage-controlled ligand presentation, surface charge or conformational changes | Protein-scale (~nm) up to cm; min-days | Reversible, non-genetic control of adhesion and downstream signaling | Hard to distinguish electrical vs mechanical and chemical effects, needs controlled electrochemistry | 16, 31 |
| Nanopipette SICM/SECM | Local ion flux, conductance, redox currents; topography | Local delivery, perturbation via applied potentials | ~10-100 nm; ms-min | High spatial resolution, minimally invasive mapping | Low throughput, operator complexity | 32 |
| Solid-state nanopores, nanopipettes, nanopores | Single-molecule ionic current signatures (DNA/RNA/proteins) | Electrically gated transport, controlled capture and translocation | nm; μs-s | Molecular selectivity, label-free | Not a direct $V_{mem}$ sensor, surface fouling, interpretation complexity | 7, 33 |
| Nanotweezers and nanobiopsy tools | Subcellular molecular content (RNA/protein/organelles) as electrical-state correlates | DEP-based capture, longitudinal sampling | ~10-100 nm tip; min-hours | Links electrical state to molecular phenotype in the same living cell over time | Specialized fabrication, low throughput, requires careful viability controls | 36, 40 |
| *Optical voltage imaging and optics-based electrical mapping* | VSDs and GEVIs, plasmonic and NV-based electrical sensors | Optogenetic actuation, optical readout enables closed loop | Subcellular-tissue; ms-hours | High content, can map network-level dynamics | Phototoxicity and bleaching (dyes), genetic manipulation (GEVIs, optogenetics), quantification challenges | 10, 41, 42 |

## 2.4. Quantum bioelectronics

The convergence of quantum mechanics and bioelectronics has also enabled novel insights into biological processes and therapeutic strategies.[8] Quantum mechanical phenomena (e.g. superposition, coherence, entanglement, and tunnelling) are being explored for their potential to enhance cancer diagnostics and treatments,[8] as well as merging electromagnetic fields (EMFs) and optics with quantum physics to precisely tune biological processes[43]. Nanoprobes and bionanoantennae, which receive signals and transduce them into biological actuators, harness EMF interactions to modulate cellular behavior at the quantum scale.[43] For example, recent studies have demonstrated that quantum biological electron tunnelling (QBET) can be sensed and actuated, affecting cancer cells specifically due to endogenous differences compared with controls. One of the earliest examples is the application of quantum dots (QDs) for cancer treatment, operating via controlled photodynamic therapy, inducing the production of reactive oxygen species (ROS) and photothermal effects.[44] Additionally, bipolar nanoelectrodes have been employed as nanoprobes, such as carbon nanotubes (CNTs), to modulate cellular electrochemistry at unprecedentedly low voltages.[45] However, precise control of quantum phenomena must rely on capacitance and underlying quantum events, such as the density of states, charge distribution, and quantum coherence.[46] To achieve precise control, it is essential to determine which specific quantum effects are at play in each system. A primary challenge is scalability and reproducibility in nanoparticle fabrication, as many synthesis methods (e.g., molecular self-assembly, electron beam lithography, and atomic layer deposition) remain costly and low yield.[43] Another challenge revolves around biocompatibility, stability, and biodistribution, which restrict their *in vivo* use.[43] Charge transfer mechanisms and quantum coherence are difficult to control in biological environments, where decoherence effects rapidly destroy quantum states.[46] Additionally, bioelectrical modulation remains poorly understood, as the precise mechanisms by which electromagnetic fields influence cellular responses at the quantum level remain elusive, making it difficult to achieve targeted therapeutic effects.[46] Lastly, immune responses tend to eliminate nanostructures, reducing therapeutic efficacy.[8] Overcoming these barriers requires advances in fabrication, surface engineering, and biocompatibility to unlock the full potential of quantum nanomedicine.

### 2.5. Optical approaches

Optical methods have significantly advanced our understanding of bioelectrical phenomena in excitable cells. Techniques such as voltage-sensitive dyes[10], calcium probes[47], and genetically encoded voltage indicators[48] (GEVIs) have enabled the monitoring of the spatial and temporal patterns of membrane potential in excitable and, more recently, in non-excitable cells[41]. However, these fluorescence-based methods are hindered by limitations such as photobleaching, which restricts long-term measurements and the scope of research questions. This limitation has driven the development of alternative more stable approaches for tracking bioelectrical dynamics. For instance, nitrogen-vacancy (NV) defects in diamonds offer a photostable charge-modulated fluorescence with experimentally validated voltage sensitivity at metal-electrolyte interfaces[49]. Similar sensors have demonstrated the ability to detect action potential in squid[50] and mouse[51] neurons. Plasmonic sensors offer another label-free approach for voltage sensing that has been successfully applied to neurons[52], cardiac cells[42], and pancreatic beta cells[41]. Beyond their label-free capability, plasmonic sensors provide several advantages including the ability to detect subcellular signaling events while simultaneously exposing how they integrate to orchestrate network-level patterns[41]. A recent study has introduced a surface plasmonic resonance microscopy (SPRM) method for studying cell bioelectrical networks monitoring connectivity in pancreatic beta cells[41]. This approach could capture coordinated electric activity, record subcellular signals that integrate to cell-level patterns and reveal the propagation of extracellular signals beyond the immediate cell-sensor interface[41]. These label-free and information-rich imaging techniques hold considerable promise for bioelectricity research. They could be explored, for instance, to further our understanding of electrical dynamics in cancer-cell network, elucidate the influence of electric fields in wound healing[53] and unfold their role in developmental pattern formation. Moreover, the ability to perform long-term imaging of electrical signaling is particularly advantageous for studying processes such as gene expression, aging, and their modulation via bioelectrical interfacing.

## 3. Prospective research areas

Emerging evidence suggests that precise mapping of cellular bioelectricity may reveal early diagnostic biomarkers, disease mechanisms and novel therapeutic targets (**Figure 2**).[2, 3, 43] In cancer, dysregulated electrical properties of tumor cells have been correlated with proliferative and metastatic behaviors.[3, 11] Similarly, age-related changes in cellular communication via altered electrophysiological patterns are increasingly recognized as contributing factors in the morphological and functional decline of tissues.[4] Controlled modulation of cellular electrical states is also gaining attention as a potential non-invasive method to induce phenotypic transitions.[2] We hereby address three envisioned areas of application, namely cancer, aging, and gene expression modulation.

### 3.1. Cancer biology, diagnostics, and therapeutics

Cancer cells exhibit a characteristic bioelectric signature (**Figure 2a**) marked by membrane depolarization, with resting potentials ranging from −10 to −40 mV compared to the −70 to −90 mV observed in healthy cells, resulting from alterations in ion channel expression and function, documented across breast[9], prostate[54], and lung[55] cancers, particularly downregulation of $K^+$ channels[56] (Kv1.3, Kv1.5, and Kv10.1), and upregulation of voltage-gated $Na^+$ channels[11] (Nav1.5, Nav1.7) and $Ca^{2+}$ channels[57] (Cav3.1, Cav3.2). This contributes to cancer pathophysiology through several mechanisms that influence fundamental cellular processes. Depolarized $V_{mem}$ affects $Ca^{2+}$ dynamics by altering the electrochemical gradient for $Ca^{2+}$ influx, leading to disrupted intracellular signaling that impacts proliferation and migration[11]. Furthermore, $V_{mem}$ influences voltage-sensitive enzymes and transporters modulating $Na^+$-dependent nutrient transporters that affect cellular metabolism.[10] It also disrupts intercellular communication through gap junctions, which are critical for tissue homeostasis.[8] Current technologies have significant limitations for studying bioelectricity in cancer cells. Traditional MEAs are optimized for detecting high-frequency action potentials in excitable cells, with challenges arising in consistently measuring the tinier voltage changes characteristic of some cancer cells.[3] In this context, the cell-material interface and electrode signal-to-noise ratio are pivotal[17], potentially enabling the recording and stimulation of cancer cells and tissues (**Figure 3a**). Voltage-sensitive dyes have captured electrical dynamics in cancer, yielding single cell-resolved measurements, but suffer from phototoxicity, limited dynamic range, and delayed timescales.[10] Preliminary *in vitro* implementations could adapt impedance-based cellular monitoring systems that have successfully detected bioelectric properties in non-excitable tissues, modified for cancer cell lines known to exhibit depolarized membrane potentials, such as MDA-MB-231/468 breast cancer cells or PC-3 prostate cancer cells.[9] The development of reliable in vitro models will enable comparisons across platforms and broader adoption towards future translation[58].

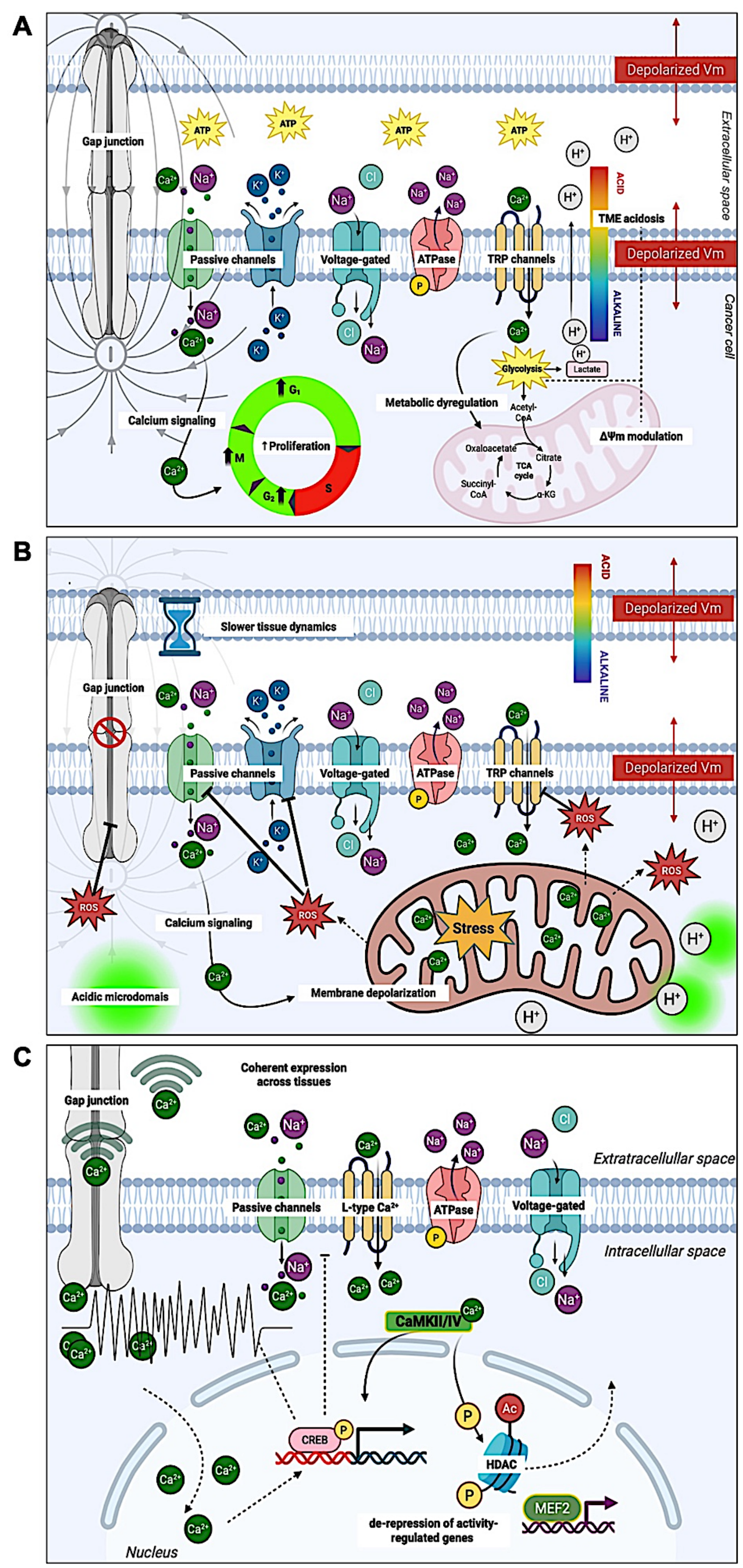

**Figure 2. Bioelectric mechanisms in cancer, aging, and gene regulation.** (A) Schematic comparison of healthy and cancer cells, showing that the latter adopt depolarized resting

potentials through reduced $K^+$ conductance and elevated $Na^+/Ca^{2+}$ entry, which favor proliferation and invasion. **(B)** Illustration of age-related electrical remodeling, where loss of hyperpolarizing currents and increased depolarizing channel activity lead to $Ca^{2+}$ overload, mitochondrial dysfunction, and senescence phenotypes. **(C)** Diagram highlighting how resting potential shifts control gene expression: depolarization drives nuclear $Ca^{2+}$ signaling and activation of transcription factors, whereas hyperpolarization favors chromatin stability and longevity-associated programs.

Key open questions remain on how field strength, duration, and waveform act. Applied field amplitudes ranging from 1-100 V/m have been shown to influence cell behavior in other contexts without causing electroporation or thermal damage, though would require longer durations (minutes to hours) in cancer, in contrast to the brief pulses used in neural stimulation.[3, 43] This would allow to determine if hyperpolarization via applied fields can reduce proliferation rates by measuring Ki-67 expression, or promote differentiation, assessed by cell-type specific markers, in cancer cells.[59] Translating optogenetic nanoswitches to cancer studies presents other technical challenges. Although standard optogenetic tools are shown to control $V_{mem}$ in neurons, cancer translation would require engineering stable expression of modified light-sensitive $K^+$ channels, which could provide a starting platform.[10] Several classes of photoswitchable $K^+$ channel modulators have been developed, including azobenzene-based compounds that change configuration upon light stimulation, altering their binding to target channels[60]. These could be screened for their ability to activate $K^+$ channels overexpressed in specific cancer types, such as Kv10.1 in breast cancer[56] or Kv1.3 in lymphomas[61]. Previous studies have demonstrated that pharmacological hyperpolarization of cancer cells through $K^+$ channel activators reduces proliferation rates and decreases migration in scratch assays.[43] Standard cell biology coupled to bioelectronics approaches could reveal mechanisms connecting electric states to molecular pathways. For example, RNA-seq analysis before and after bioelectric normalization could identify voltage-responsive gene networks, and metabolic flux analysis could determine how $V_{mem}$ shifts affect the Warburg effect characteristic of cancer metabolism.[62]

### 3.2. Probing and modulating cellular aging

The bioelectric dimension of senescence is starting to be considered as a set of parameters that can be probed to induce anti-aging effects (**Figure 2b**).[4] A hallmark of aging cells is a depolarized resting membrane potential (~ –50 mV) compared to youthful cells (~ –70 mV), arising from ion channel remodeling.[4] Key hyperpolarizing potassium currents weaken with age, for instance expression and activity of $K^+$ channels (decline in Kv1.3, Kv1.5 and KCa3.1), and depolarizing currents increase via upregulated voltage-gated $Na^+$ and TRP cation channels, which drive aging phenotypes by aberrantly activating voltage-sensitive enzymes.[63] Depolarization triggers phospholipases and phosphatases that alter signaling cascades governing cell senescence, including pro-inflammatory pathways that underlie the senescence-associated secretory phenotype (SASP).[64] Depolarization also boosts intracellular $Ca^{2+}$ levels by opening voltage-gated $Ca^{2+}$ channels, with consequent $Ca^{2+}$ release leading to mitochondrial overload, metabolic impairment, and elevated reactive oxygen species (ROS).[65] These events back into gene regulation: $Ca^{2+}$-dependent activation of histone deacetylases and DNA methyltransferases in depolarized cells drives chromatin, hence cell shape and morphology, toward pro-aging configurations. Generally, more hyperpolarized (negative) $V_{mem}$ correlates with enhanced cellular longevity, whereas chronic depolarization accelerates aging processes.[66] Despite these insights, the electrical aspect of aging remains comparatively unexplored, partly due to technical gaps in measuring and manipulating subtle bioelectric signals.[3, 43] In this context, molecular nanoswitches such as optogenetic ion pumps might provide cell-type specific voltage control, and light-driven proton pumps (Arch) or chloride pumps (halorhodopsin) could hyperpolarize aged cells, counteracting depolarization, whereas depolarizing channels like channelrhodopsins could be deployed to dissect how brief pulses of depolarization might trigger senescent phenotypes.[67] Early studies already hint at the therapeutic potential of tuning bioelectricity in aging. In C. elegans, genetic enhancements of $K^+$ channel activity (thus preventing depolarization) have shown to extend lifespan, and pharmacologically hyperpolarizing mammalian cells delayed senescence.[68] In Drosophila, hyperpolarization was shown to slow systemic aging via cell-nonautonomous signaling, demonstrating that local electric changes propagate organism-wide.[69] Bioelectronic devices could be deployed across specific contexts where senescence manifests differently. In neurons, for example, gradual depolarization disrupts $Ca^{2+}$ homeostasis and synaptic plasticity[70]; in cardiomyocytes, impaired $K^+$ channel activity contributes to reduced contractility[71]; in fibroblasts, depolarization accelerates transition to senescence-associated secretory phenotypes[72]. Recording progressive $V_{mem}$ drift over weeks in organoid or cell cultures could map tissue-specific senescence processes, enabling longitudinal analysis and current injection to restore hyperpolarized states.

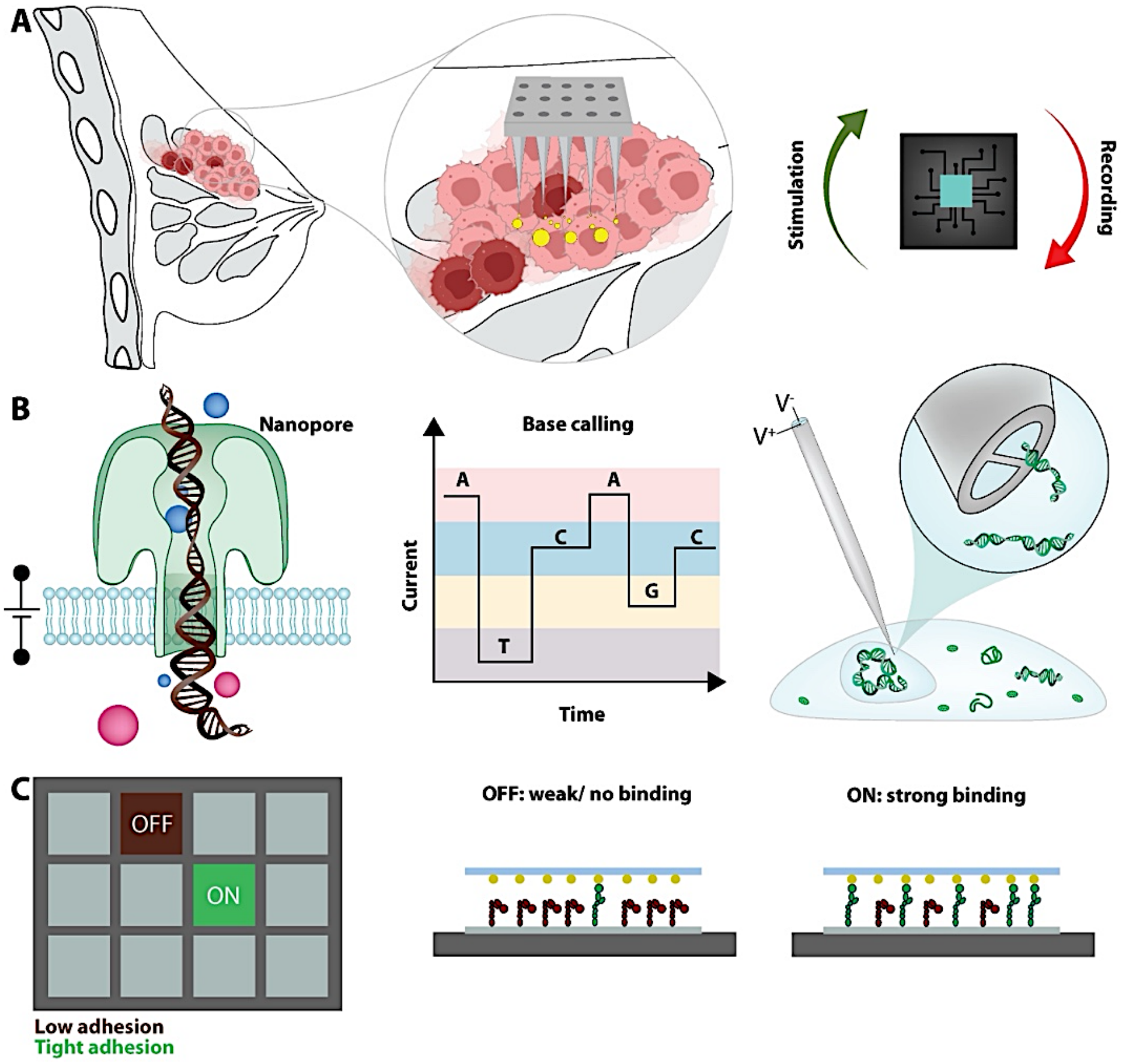

**Figure 3. Envisioned applications of bioelectronic platforms in oncology, aging, and gene expression modulation. (A)** Concept for probing tumor bioelectricity: arrays of penetrating electrodes designed for low noise and optimal cell-material adhesion record depolarized states within cancerous tissue while delivering controlled stimulation, allowing dynamic readout of malignant electrical signatures and assessment of how imposed fields may suppress tumor activity. **(B)** Single-molecule tools such as nanopores and nanotweezers can provide access to ionic currents from individual membrane channels, $Ca^{2+}$ imbalance, as well as mitochondrial stress and ROS production, potentially providing a window into electrical events associated to senescence. By extracting or reading nucleic acids and proteins from single cells, nanotweezers

could track senescence-associated states and test strategies to restore hyperpolarized potentials. **(C)** Electroactive switchable interfaces yield surfaces whose adhesive properties change in response to applied voltage. By toggling between non-binding and high-affinity states, these can regulate integrin engagement, cytoskeletal architecture, and downstream nuclear organization, reversibly tuning gene expression and transcriptional activity.

This closed-loop approach provides a foundation for developing bioelectronic manipulation protocols, in which the parameters of stimulating electric currents are precisely aligned with the targeted membrane potential and the corresponding cell function. Nanopores could detect redox-active metabolites or tracking ion fluxes through individual channels that become dysregulated with age[33] (Figure 3b). Nanotweezers, by contrast, could extract and reposition mitochondria or cytoplasmic fractions from single cells in an aging population, linking bioelectric depolarization with organelle-specific dysfunction such as loss of mitochondrial membrane potential or altered ER-mitochondria calcium exchange[36]. Complementary readouts could include impedance spectroscopy to track extracellular matrix stiffening, as stiffness is tightly coupled to both senescence and altered electrical states[73].

### 3.3. Gene expression reprogramming

Gene expression is strongly influenced by the electrical state of cells, with $V_{mem}$ shifts acting as a primary layer of regulation (**Figure 2c**).[13] In non-excitable cells, changes in resting potential from hyperpolarized (~ –90 mV) to depolarized (~ –20 mV) values shape transcriptional outcomes by modulating signaling cascades, transcription factor activity, and chromatin architecture.[74] Depolarization promotes the opening of voltage-gated $Ca^{2+}$ channels, leading to $Ca^{2+}$ influx and activation of calcineurin and CaMK pathways that drive nuclear translocation of NFAT, CREB, and NF-κB, thereby altering the expression of hundreds of genes involved in metabolism, stress adaptation, and lineage specification.[74] Voltage-sensitive phosphatases (VSPs) can transduce membrane depolarization into changes in phosphoinositide levels, which may disrupt PI3K/Akt signaling.[75] On the other hand, hyperpolarization has been shown in neural progenitors to restrict Wnt/β-catenin signaling via altered β-catenin localization, and inhibition of KCNQ1 channels (causing depolarization) correlates with downregulated Wnt activity via β-catenin mislocalization.[76] Electrical shifts are not confined to single cells: gap junctional coupling allows $V_{mem}$ changes to propagate across tissues, producing domains of synchronized transcriptional activity that underlie morphogenesis and coordinated differentiation.[2] A further layer of control arises from the spatial compartmentalization of electrically driven signaling. Nuclear $Ca^{2+}$ transients, generated when depolarization-induced cytosolic $Ca^{2+}$ influx couples to nuclear pore complexes and inner nuclear envelope channels, exert particularly strong effects on transcriptional regulation by activating CREB- and NFAT-dependent promoters directly within the nucleus.[14] Nuclear-restricted $Ca^{2+}$ events differ from global cytoplasmic signals in duration and amplitude, producing transcriptional specificity that could be dissected with low-noise microelectrodes configured to deliver focal depolarizing currents near the perinuclear region.[14] In addition to $Ca^{2+}$, $V_{mem}$ shifts modulate redox balance and pH, both of which target chromatin-modifying enzymes, creating voltage-sensitive epigenetic checkpoints.[76]

Applied bioelectric fields could therefore be tuned not only to activate acute transcription factors but also to imprint long-lasting transcriptional programs through chromatin modifications.[76] At the tissue level, engineered electrical gradients imposed across organoids have been shown to bias regional expression of positional identity genes, suggesting that bioelectrical patterning can recapitulate aspects of morphogenesis.[77] Nanopores, providing both gated transport and molecular sensing functions,[7, 33, 78] could be capable of monitoring RNA or protein markers of transcription in real time as cells respond to electrical cues, and extracting subcellular samples from living cells without inducing cell death might allow measurement of chromatin modifications as a response to localized stimuli. Electroactive switchable surfaces might introduce another level of control by exposing or hiding integrin-binding motifs under electrical stimulation,[16] thereby engaging the cytoskeleton and transmitting mechanical forces to the nucleus, where they reorganize lamina structure and chromatin state to reinforce or redirect transcriptional programs (**Figure 3c**). Electrical stimulation has already been shown to induce expression of developmental regulators such as SOX9 in chondrogenic contexts, TBX5 in cardiac differentiation, and HOX clusters that encode positional identity. In cultured stem cells, brief pulses of current induce lineage-defining transcription factors.[79] Advances in biology have started to create the experimental conditions where gene networks could be electrically interrogated and modulated in real time without genetic intervention.

## 4. Outlook

Moving from high-precision measurements to practical bioelectronic tools for oncology and aging will require the field to treat bioelectricity as a set of variables that can be quantified. For non-excitable cells, this shift is particularly urgent because the signals of interest are mostly smaller, slower, and more context-dependent than action potentials, and they unfold within heterogeneous microenvironments (ECM remodeling, hypoxia, inflammation) that can dominate over the specific variable of interest. Progress therefore depends on coupling recording with stimulation, and on designing experiments that convert correlations (e.g., depolarized $V_{mem}$ in tumors or senescence) to relationships between electrical inputs and molecular outputs. A first practical step is to define standardized electrical endpoints that can be compared across devices and biological models. In addition to reporting mean $V_{mem}$, non-excitable systems will likely require distributions and spatial structure (e.g., gradients, network effects and temporal scales) because tissue-level phenotypes are expected to emerge from multicellular states. Other variables that can be measured with electrical tools such as impedance (barrier integrity, cell-substrate coupling), and electrochemical markers (local pH, redox) can provide additional readouts useful for interpretation. On the device side, establishing a standard on working electrode materials and geometries, impedance, stimulation numbers, expected waveforms, estimated field exposure, temperature, and biological conditions, would help improving interpretability and reproducibility, just as how stimulation became standardized in neurotechnology.

Device translation for non-excitable cell biology will also require device stability for long experiments, as many of the envisioned interventions in cancer, aging, and gene regulation described herein happen in minutes, hours or days. Materials that have transformed neural and cardiac interfacing such as TiN, conducting polymer coatings, and mixed ionic-electronic conductors in transistors, offer promising routes to reduce impedance and operate at low voltages, but their use in non-excitable contexts has yet to be demonstrated. Next, progress will depend on field delivery in 3D constructs, such as tissue slices, spheroids and organoids. Equally important is where these technologies are validated. To aim at clinical relevance, electrical interfaces should be tested in models that preserve multicellular coupling and microenvironmental constraints while remaining compatible with early testing in this initial phase. Here, the most impactful near-term experiments are longitudinal studies for mapping slow drift in electrical state and impact of electrical modulation to map, maintain or restore a target voltage range while tracking phenotypic outcomes (e.g., invasion, differentiation). The value of single-cell tools discussed herein is that they can connect electrical manipulations to molecular events by enabling controlled sampling of RNA, protein, or organelles from the same living cells, reducing the reliance on external assays. From a molecular biology standpoint, the next phase should target experiments that link electrical inputs to known transducers highlighted in the previous sections, such as calcium channel dynamics, voltage-sensitive enzymes, gap-junctions, and chromatin remodeling. Electroactive switchable interfaces are particularly valuable here because they can reversibly impose controlled changes at the interface (e.g., integrin engagement and cytoskeletal organization), enabling to distinguish between material interface effects and biological voltage shifts.

Once the engineering and scientific foundations are more solid, translation could follow examples that have already navigated safety, scalability, reliability and broader adoption. Bioelectronic medicine so far has been defined around dosing, device performance, and medical regulation, providing a possible path to building solutions suitable for patient translation. In the nearer term, some of the most realistic translational opportunities may arise in in vitro early diagnostics, drug screening, and electricity-based drug development across the pharmaceutical sector.